\documentclass[preprint]{aastex63}
\usepackage[utf8]{inputenc}

\usepackage{amsmath}
\usepackage{amssymb}

\graphicspath{{./}{figures/}}

\begin{document}

\title{Cloud Continents: Terraforming Venus Efficiently by Means of a Floating Artificial Surface}

\author[0000-0002-4884-7150]{Alex R. Howe}
\address{NASA Goddard Space Flight Center}
\email{alex.r.howe@nasa.gov}

\begin{abstract}

The similarity of Venus and Earth in bulk properties make Venus an appealing target for future colonization. Several proposals have been put forward for colonizing and even terraforming Venus despite the extreme conditions on the planet's surface. Such a terraforming project would face large challenges centered around removing Venus's massive carbon dioxide atmosphere and replacing it with a habitable environment. I review past proposals and propose a new method for terraforming Venus by building an artificial surface in the much more hospitable upper atmosphere where the temperature and pressure are both Earth-like. Such a surface could be built with locally produced materials and would float above the bulk of the atmosphere using nitrogen as a lifting gas. This would allow the engineering of a breathable atmosphere above the surface and would remove the need to import or export extreme amounts of mass, except for comparatively modest quantities of water. The engineering, logistical, and energy requirements of this method are surveyed. I find that such a terraforming project could be completed in a minimum of 200 years in a best-case scenario, comparable to other proposals, with significantly lower resource costs.

\end{abstract}








\section{Introduction} \label{sec:intro}

The terraforming of Mars to produce a habitable environment on the surface is a frequent topic of discussion in long-term human spaceflight (e.g. \cite{Zubrin,MarsReview,Mars18a,Mars18b}). The reason for this is clear: the surface conditions on Mars are the most similar to those on Earth of any body in our solar system and thus would require the least modification to adapt its surface for human habitation. Venus, in contrast, appears to be a much less likely candidate for terraforming due to its extremely thick atmosphere and a very high surface temperature produced by a runaway greenhouse effect. Nonetheless, the potential terraforming of Venus has received some consideration, beginning with Sagan \cite{Sagan}. Venus also provides some advantages over Mars for colonization with its near-Earth-like surface gravity, an atmosphere thick enough to provide robust protection from cosmic rays \cite{Zeitlin} and UV radiation \cite{Mogul} compared with Mars, and shorter travel time from Earth.

The challenges for terraforming Venus's surface are great, requiring a significantly higher degree of speculative planetary engineering than Mars. The main requirements for producing a habitable environment on the surface are:
\begin{itemize}
    \item Removing or sequestering the $\sim$90 bars of carbon dioxide from the atmosphere.
    \item Producing a breathable nitrogen-oxygen atmosphere.
    \item Cooling Venus from its current surface temperature of 735 K.
    \item Replacing the planet's lost water.
    \item Speeding up the planet's rotation from its current value of a solar day lasting 117 Earth days. (However, this is probably not essential for habitability if the albedo can be artificially controlled, cf. \cite{Tidal}).
\end{itemize}

An alternative to terraforming the surface of Venus would be to colonize the upper atmosphere \citep{Landis}. Taking advantage of the twin facts that the upper atmosphere of Venus at an altitude of 50-55 km has a similar temperature and pressure to Earth's surface, and that nitrogen and oxygen are lifting gases in Venus's CO$_2$ atmosphere, it would be possible to construct ``cloud cities'' using aerostats filled with breathable air that could be inhabited by humans. This method sidesteps most of the problems of colonizing the surface; it would be achievable with current technology, and it has reached the level of a mission concept \citep{HAVOC}.

In this paper, I propose a method of terraforming Venus that similarly sidesteps most of the problems posed by the hostile conditions on the surface by expanding the cloud city concept to ``cloud continents,'' building an entire artificial surface floating in the upper atmosphere. While still speculative, this could be done far more efficiently than other proposed methods for terraforming Venus's surface while allowing a far more robust and expansive habitable environment than could be achieved by conventional aerostats. Nearly all of the requisite materials for the project would be produced locally on Venus, while water could be delivered from Mars via space elevator. If the requisite technology were sufficiently developed, the process could plausibly be energy-limited, and if all available solar energy resources are exploited at currently available efficiencies, it could potentially be completed in as little as 200 years, comparable to the fastest timescales for other proposed methods \citep{Birch}.

Certainly, this estimate of 200 years is an extreme best-case scenario and would require substantial development of technology to be possible at all. All of the materials needed for such an effort have been demonstrated in the laboratory, but cannot yet be manufactured on an industrial, let alone planetary scale. The geoengineering techniques involved would be immensely more efficient on Earth (or even Mars), so it would likely be a long time before we had reason to turn them to Venus rather than applying them at home. However, the goal of this paper is to provide an outline for how the terraforming of Venus could be accomplished using foreseeable technology on something close to a human timescale, and to demonstrate that it is a more tractable problem than is usually assumed.

This paper is organized as follows. In Section \ref{sec:literature}, I give an overview of past proposals for terraforming Venus and their limitations. Section \ref{sec:design} outlines the design and construction process for the artificial surface. Section \ref{sec:env} explores options for building an Earth-like environment on the new surface. Section \ref{sec:costs} examines the costs of construction, the logistics of importing water, and a potential timetable. I summarize my conclusions in Section \ref{sec:conclusion}.

\section{Previous Proposals} \label{sec:literature}

The first major proposal for terraforming Venus was made by Sagan \cite{Sagan}, even before the extreme surface conditions were known. At that time, a ``carboniferous swamp'' was considered to be one of several possible models for the surface environment, and microwave emissions had only raised the possibility of a strong greenhouse environment in excess of 600 K. However, it \textit{was} known that Venus's atmosphere was primarily made of CO$_2$, which would need to be removed to make the planet habitable for humans. Sagan suggested that this could be achieved by seeding the cloud layer with photosynthetic algae to break down the CO$_2$ into oxygen and organic material. However, as we understand Venus today, it is too dry and its atmosphere too thick for changing the atmospheric chemistry alone to work; even if the atmospheric composition were successfully changed, the problem would remain of removing the tens of bars of oxygen that would be produced by photosynthetically breaking down all of the CO$_2$.

Adelman \cite{Adelman} suggested colliding asteroids with Venus to strip it of its excess atmosphere. This would be a difficult process, as the mass of the impactors would need to be greater than the mass of the atmosphere itself \citep{MassLoss}, which is on the order of $5\times10^{20}$ kg. Like many such proposals, this carries the implicit challenge of moving a dwarf planet's worth of mass across interplanetary distances, itself a very expensive process.

Birch \cite{Birch} lays out a detailed program for terraforming Venus with the claim that it could theoretically be accomplished in as little as 200 years. Most of this time would be needed to cool the planet by placing a large, space-based sunshade in front of Venus to block all sunlight from reaching it and waiting for the atmosphere to condense and freeze. The frozen atmosphere could then be artificially sequestered by a thermally-insulating cover. However, long-term, the CO$_2$ would need to be exported off the planet. Meanwhile, a water ocean would be provided by breaking up Enceladus and delivering it in pieces to the planet. This also has the problem of needing to move large amounts of mass, and even if Enceladus were replaced with a different body, we may be reluctant to scuttle an entire large moon for the project. Birch does raise the possibility of constructing an artificial surface on Venus, but considers creating a ``natural'' geosphere to be more desirable.

Bullock \& Grinspoon \cite{Bullock}, though not addressing terraforming directly, suggest that it may be possible to sequester the CO$_2$ in Venus's atmosphere by chemically reacting it with calcium minerals on the surface. This would remove the need to export the atmosphere, but the natural timescale for this process is estimated at 10$^7$--10$^8$ years, so speeding it up to a practical timescale would be challenging.

More recently, Landis \cite{Landis} surveyed the various proposals for terraforming Venus and found colonizing the upper atmosphere to be much more feasible. This would most naturally be done with aerostats, but he also suggests building foundations up from the surface and using genetically-modified organisms to produce a breathable atmosphere in the cloud layer. However, building up 50 km from the surface presents significant engineering challenges, and it would be necessary to engineer the whole atmosphere at once due to vertical mixing. An artificial surface dividing the upper and lower layers of the atmosphere would be needed to limit the terraforming requirements to the habitable-pressure zone.

\section{Surface Design} \label{sec:design}

An artificial surface on Venus must perform two tasks: (1) reliably separate the bulk of the CO$_2$ atmosphere below it from the eventual oxygen-nitrogen atmosphere above and (2) support its own weight and any colonies built on it by buoyant force both during and after construction. (For a fully enclosed surface, the supporting force is the uniform upward pressure of the atmosphere rather than buoyancy, but as there are several advantages to maintaining ``neutral buoyancy'' of the finished surface as discussed below, both cases are described using buoyant forces for simplicity.) The first requirement means that the planet's surface must be fully enclosed before the atmospheric engineering can take place. This requires a large initial investment, but, unlike most other proposals, it is the only part of the terraforming process that is dependent on a single, large step, whereas the rest of the operations can be done piecemeal.

The second requirement means that the artificial surface should be neutrally buoyant. While a fully enclosing surface could be heavier and still be supported by gas pressure once completed, neutral buoyancy will be essential during construction, and it would offer a more stable platform for the uneven weight distribution of a partially built-up surface. A neutrally buoyant surface would also decrease the rate of leakage from the lower atmosphere in case of damage, thus reducing maintenance costs.

These considerations suggest an artificial surface built in two parts: first, a lightweight surface enclosing the entire planet that need only support its own weight and be resistant to tearing. Second, larger floating ``islands'' and ``continents'' could be constructed embedded in the surface with enough lifting power to support full-scale colonies with a robust infrastructure including housing, manufacturing capacity, croplands, and an imported biosphere.

Both parts of the surface could be constructed with local materials, using solar power to break CO$_2$ into carbon and oxygen. The carbon would be used to manufacture carbon nanotubes or similar structures for construction material, while the oxygen could be used as a lifting gas or stored for future use. The 3.5 bars of nitrogen already present in the atmosphere would be the primary lifting gas. Any metals or other minerals needed for construction could be mined robotically from the ground level. Additionally, both parts of the surface could be constructed in parallel with each other, and settling the planet could occur at the same time through enclosed, ``conventional'' aerostat colonies.

\subsection{Enclosing the Surface}

The first step of the terraforming process will be to fully enclose Venus with a lightweight surface at an altitude of $\sim$50 km to separate the two parts of the atmosphere. The main structural limitation at this stage will be wind forces. In Venus's atmosphere, zonal wind speeds at the 1-bar level are 40-60 m/s \citep{VenusWind,VenusAtmos}, indicating that the surface must withstand horizontal wind shear of tens of meters per second. Both meridional and vertical winds are significantly weaker at $\lesssim$5 m/s, although vertical winds due to convection will be a source of unwanted mixing of the upper and lower atmosphere through any holes in the surface.

A surface made as a single piece will not be able to withstand these wind forces over thousands of kilometers. Instead, we may envision large, interlocking hexagonal tiles, a few tens of meters thick, with the joints between them providing the slack to resist the wind forces. These tiles would only need to support their own weight, so they could be manufactured with similar materials and techniques to conventional airships (perhaps using carbon fiber if sufficient nanotube production is not yet available). The top surface of the tiles could be mirrored to maintain the planet's Bond albedo of 0.76, which will help with later climate control. Tears in the surface could be repaired by replacing damaged tiles. For tiles 100 meters in width diagonally, comparable to a large airship, $7.2\times 10^{10}$ of them would be sufficient to enclose the planet at an altitude of 50 km.

Maintenance of this tile surface would be essential to maintaining the chemical composition of the upper atmosphere, as any holes would allow CO$_2$ to leak from below. A neutrally buoyant surface would produce the least leakage, but it could still be significant in severe cases. Assuming Earth-like conditions for the upper atmosphere and a 5 m/s vertical flow, a tear with an area of 1 km$^2$ would leak CO$_2$ at a rate of $8.04\times10^{11}$ kg/day. This rate of leakage would increase the CO$_2$ concentration in the upper atmosphere by 0.101 ppm/day until the tear is closed. Adequate maintenance systems would need to detect and deflect any asteroids capable of causing tears this large before they reach the planet, while being able to replace the tiles fast enough to keep up with lesser damage. If limited to aircraft speeds, it may take several days to dispatch repair equipment to remote areas, so it may be preferable to build a surplus of replacement tiles and station them at depots around the planet.

\subsection{Floating Landmasses}

Building floating landmasses large enough to support cities, infrastructure, and ecosystems similar to those on Earth requires more careful consideration. These landmasses will be limited by the inventory of chemicals available on Venus and particularly in the atmosphere for construction. The inventory of atmospheric species of greater than 1 ppmv abundance (based on surface abundances) is shown in Table \ref{tab:inventory}. For this purpose, I will assume a surface pressure of 92.0 bars. Clearly, CO$_2$ and N$_2$ are the species of interest, and the problem of removing the sulfur dioxide (and indeed, sulfuric acid) from the upper atmosphere will be secondary to changing it to a nitrogen-oxygen composition.

\begin{table}[htb]
    \centering
    \begin{tabular}{l | r | r}
    \hline
    Species  & Mole Fraction & Areal Density  \\
             & \          & (kg m$^{-2}$) \\
    \hline
    CO$_2$   &  96.5\%       & 1027000     \\
    N$_2$    &   3.5\%       &   23710     \\
    SO$_2$   & 200 ppm       &     310     \\
    Ar       &  70 ppm       &      67.8   \\
    COS      &  55 ppm       &      79.8   \\
    H$_2$O   &  30 ppm       &      13.1   \\
    CO       & 12-40 ppm     &    8-27     \\
    He       &  12 ppm       &       1.16  \\
    Ne       &   7 ppm       &       3.39  \\
    Total    & 100.0\%       & 1051000     \\
    \hline
    \end{tabular}
    \caption{Chemical inventory of Venus's atmosphere, assuming a surface pressure of 92.0 bar. (Based on surface abundances.) \citep{Basilevsky}}
    \label{tab:inventory}
\end{table}

The floating landmasses would be constructed with a honeycomb structure, not too dissimilar from the tiles, but much taller, kilometers high rather than tens of meters. As with the lighter tiled surface, the main construction material will be carbon nanomaterials made with carbon extracted from the atmosphere. Because carbon nanotubes are much stronger in tension than compression, it will likely be better to use a different structure such as aggregated diamond nanorods \citep{Nanorods} or some form of diamond-nanotube composite for the load-bearing walls. Diamond is reported to have a compressive strength of 250 GPa and a shear strength of 75 GPa at nano-scales \citep{Diamond}. This is comparable with the reported tensile strength of carbon nanotubes of 100 GPa \citep{CNTref} and well above the tensile strength reported for carbon nanotube bundles of 17 GPa \citep{CNTbundle}. As we will see, the stress placed on the load-bearing structures of the artificial surface will be $\sim$10.8 GPa in the most extreme case and could be made significantly lower, so it is plausible that adequate carbon nanostructures could be found.

In principle, a large fraction of the Venusian atmosphere could be converted to building materials, the limitation being how much surplus oxygen could be safely stored in the honeycomb as lifting gas before becoming a fire hazard. However, for definiteness, I will focus on the most efficient design in terms of lift-to-weight ratio, in which exactly as much CO$_2$ is spent as is needed to produce the same partial pressure of oxygen as Earth in the upper atmosphere, while still using all of the surplus nitrogen for lift. This design will require 0.32 bar of CO$_2$, and it will be able to support an additional 7440 kg/m$^2$ of mass loading.

\begin{figure}[h!]
    \centering
    \includegraphics[width=0.66\textwidth]{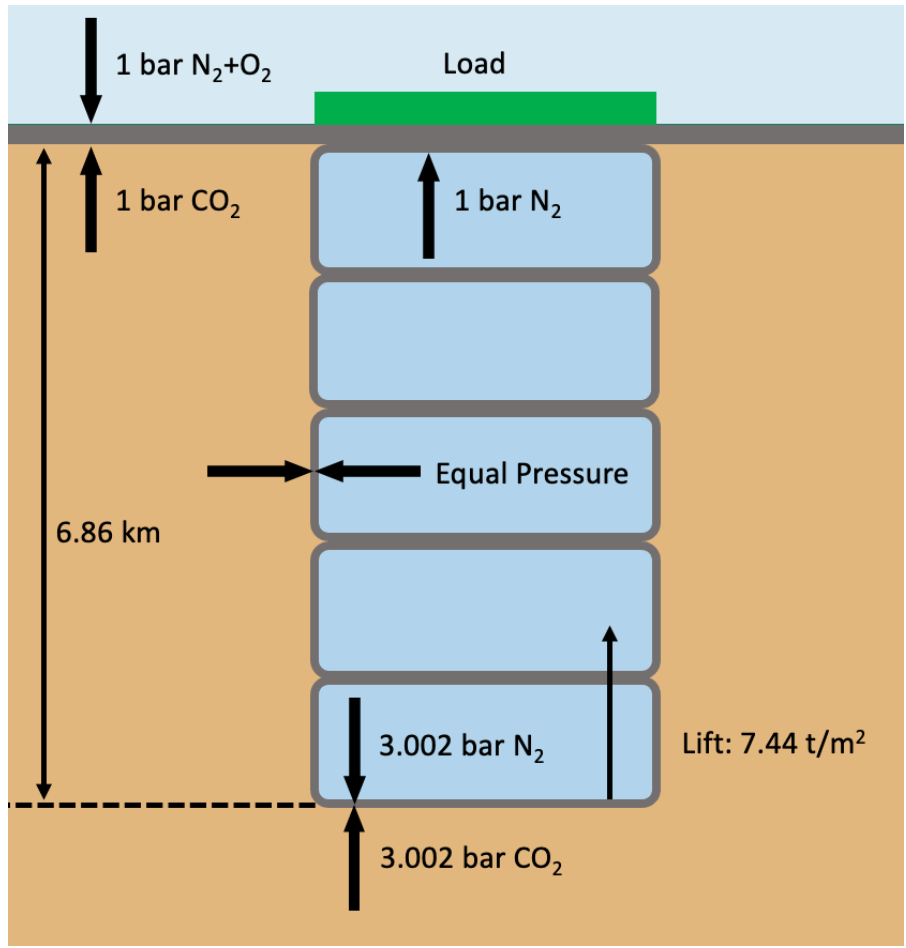}
    \caption{Illustration of the floating landmass design. The load-bearing surface is supported by a honeycomb filled with lifting gas (either nitrogen or a nitrogen-oxygen mixture) with each layer of cells pressurized to the ambient pressure to minimize stress during construction. In all designs, the honeycomb will be several kilometers tall and will support several tons of mass loading per square meter.}
    \label{fig:surface}
\end{figure}

The general design of the floating landmasses is illustrated in Figure \ref{fig:surface}. Due to the difference in scale height between N$_2$ and CO$_2$, the honeycomb should be built in layers, with each layer pressurized to the same level as the ambient atmosphere, in order to prevent a large pressure differential from developing at either the top or bottom of the column. The thickness of the honeycomb is dependent on the amount of lifting gas used, but in all cases will be several kilometers, and the load on top of the landmass must equal the lifting capacity in order to maintain neutral buoyancy.

In order to produce a pressure of 1 bar above the artificial surface, we require an areal air density of $\frac{1 bar}{g_V} = 11620$ kg/m$^2$, where $g_V$ is the gravity on Venus (noting that at altitude, the gravity will be slightly lower than at ground level: 872 cm/s$^2$ rather than 887 cm/s$^2$). To have a molar fraction of 79\% N$_2$ and 21\% O$_2$ requires this atmosphere to include 8910 kg/m$^2$ of nitrogen. This is 37.6\% of the atmosphere's nitrogen inventory. The remaining nitrogen, 14800 km/m$^2$, will be pumped into the honeycomb to serve as lifting gas. With the layers of the honeycomb pressurized to the ambient pressure, the lift is given simply by the ratio in molecular weights between N$_2$ and CO$_2$. The nitrogen will thus displace 23260 kg/m$^2$ of CO$_2$ extending down to a pressure level of 3.002 bar. Assuming an average air temperature the same as Earth's, 288 K (see Section \ref{sec:env}), the scale height of CO$_2$ will be 6240 m, and the thickness of the honeycomb will be 6860 m. (Given the temperature difference over such a change in altitude, the actual thickness may need to be several hundred meters greater.)

The buoyancy of this volume of nitrogen is 8460 kg/m$^2$, which must include the weight of the honeycomb itself plus anything built on top of it. The mass of the honeycomb is equal to the amount of carbon extracted from CO$_2$ to produce the oxygen in the atmosphere, giving it a mass of 1020 kg/m$^2$, leaving 7440 kg/m$^2$ of available lifting capacity for the floating landmass. This value also determines the strength required for the load-bearing structure, as the walls of the honeycomb must extend for its whole height. The internal structure of the honeycomb is not certain, and cross supports will be needed between the cells, if not elsewhere, but for definiteness, I will assume that two thirds of the carbon goes into the vertical walls, as would be the case for a stack of uniform cubes. In the worst-case scenario, the carbon nanostructure will have the density of diamond as opposed to lighter allotropes of carbon, $\rho = 3.515$ g/cm$^3$. In this configuration, a horizontal cross section of the honeycomb would have a fill fraction of $f = \frac{2\sigma_C}{3\rho H} = 2.82\times10^{-5}$, where $\sigma_C$ is the area density of carbon, and $H$ is the height of the honeycomb. The base of the honeycomb will be under a pressure of 3.002 bar, so the necessary compressive strength to support it will be $P/f = 10.8$ GPa.

The mechanical stresses on the honeycomb could be reduced by processing more CO$_2$ into carbon and adding the surplus oxygen to the lifting gas supply. The disproportionately greater carbon supply would allow the walls of the honeycomb to be thicker, reducing compressive stress. Alternatively, the height of the entire column could be made shorter, supporting less weight and using less nitrogen for lift. This would be an ideal solution if the full lifting capacity of the honeycomb is not needed.

Table \ref{tab:loading} provides a breakdown of the mass loading and other mechanical properties for three possible designs of floating landmasses: a ``standard'' design using the full nitrogen budget of the atmosphere, a ``heavy'' design that adds enough surplus oxygen to fill the entire honeycomb with breathable air, and a ``light'' design that provides the minimum amount of lift estimated to support a viable biosphere on the surface. Both the heavy and light designs require significantly less structural strength.

\begin{table}[htb]
    \centering
    \begin{tabular}{l | r | r | r}
    \hline
              & Standard & Heavy & Light \\
    \hline
    Mass Loading  & \  & \  & \\
    (kg m$^{-2}$) & \  & \  & \\
    \hline
    N$_2$                    & 23710 & 23710 & 16310 \\
    \hspace{12pt}Atmospheric &  8910 &  8910 &  8910 \\
    \hspace{12pt}Lifting     & 14800 & 14800 &  6000 \\
    O$_2$                    &  2710 &  7210 &  2710 \\
    \hspace{12pt}Atmospheric &  2710 &  2710 &  2710 \\
    \hspace{12pt}Lifting     &     0 &  4500 &     0 \\
    Structure                &  1020 &  2710 &  1020 \\
    Total Mass               & 27440 & 33630 & 18640 \\
    Displacement             & 23260 & 29450 &  9430 \\
    Net Lift                 &  7440 &  7440 &  2410 \\
    \hline
    Base Pressure (bar)      & 3.002 & 3.534 & 1.810 \\
    Column Height (m)        &  6860 &  7880 &  3700 \\
    Compressive Stress (GPa) &  10.8 &   5.5 &   3.5 \\
    \hline
    \end{tabular}
    \caption{Mass loading for three floating landmass designs. The ``Standard'' design uses all of the nitrogen in the atmosphere, but only as much CO$_2$ as is needed to produce a breathable nitrogen-oxygen upper atmosphere. The ``Heavy'' design uses enough CO$_2$ to use breathable air as a lifting gas throughout the honeycomb. The ``Light'' design is based on as estimate of the minimum lift needed to support a viable biosphere (including soil, water, and biomass).}
    \label{tab:loading}
\end{table}

\pagebreak

\subsection{Heat Balance of the Atmosphere}

One complication to this design is that enclosing the surface with these floating landmasses will change the temperature profile of the lower region of the atmosphere. The upper surface of the landmasses and the adjacent topmost layer of the gas-filled honeycomb would be at approximately the same temperature as the present atmosphere of Venus at the same altitude. However, the loss of solar heating to the deeper layers would cause them to cool and contract, diminishing the pressure support for the surface. The cooling rate would depend on the rate of thermal transport in Venus's atmosphere. This rate would likely be slowed by the honeycomb itself cutting off convection in the top several kilometers and may be slowed more if the honeycomb is designed to be insulating, but it would still likely not maintain a steady state over long periods.

To restore the thermal equilibrium of the lower region of the atmosphere, a simple solution would be to place ``windows'' of transparent material in the artifical surface. This could be done by building the aerostat tiles of parts of the initial enclosing surface our of transparent plastics, either imported directly, or made by reacting carbon with hydrogen extracted from the sulfuric acid clouds and the small supply of water vapor. These would then allow solar radiation to reach the lower atmosphere and maintain it in thermal equilibrium. The floating landmasses could not be extended across the windows, but if the honeycomb is sufficiently insulating, the windows would likely need to cover only a small percentage of the surface. Further research is needed to more accurately predict the thermal transport in such an enclosed atmosphere.

\pagebreak

\subsection{Parallel Colonization}

Unlike the initial enclosing of the planet with the tile surface, building the floating landmasses could be done gradually because the other steps in the terraforming process (delivery of water, soil, and building material and changing the upper atmospheric chemistry) can be done in parallel with it and do not rely on the honeycomb structure being completed. However, open-air colonies and ecosystems \textit{do} depend on the upper atmosphere being changed to an Earth-like composition. This is expected to be a protracted process requiring extracting 37.6\% of the nitrogen from the lower atmosphere and splitting all of the needed carbon dioxide to supply oxygen, as well as pumping out all of the CO$_2$ initially occurring in the upper atmosphere. Thus, these open-air colonies could be constructed only after the terraforming process is largely complete.

However, the human colonization of Venus could proceed unimpeded in parallel with these efforts. Once small (a few km$^2$) floating islands are constructed, large habitat domes could be built on top of them. Rather than freestanding domes, these habitats could be enclosed by the same aerostat tiles covering the rest of the planet, allowing much larger enclosed volumes than can be achieved with self-supporting structures. While the aerostat tiles would be floating above a lighter nitrogen-oxygen atmosphere, lift could be maintained by partially filling the tiles with helium. The partial pressure of 1.1 mbar of helium at ground level could provide an ample supply. Light could be let in by building the tiles out of the same transparent plastics as the surface windows. Eventually, once the exterior air becomes breathable, the domes would be dismantled and recycled.

\section{The Terraformed Environment} \label{sec:env}

\subsection{Temperature}

With a breathable atmosphere in place, the next question will be how to regulate the climate of the terraformed Venus. Being much closer to the Sun than Earth, a precisely Earth-like atmosphere and surface (with much lower albedo than the present Venus) will become too hot for human habitation. Many proposals suggest a sunshade to reduce the sunlight reaching the surface to Earth-like levels, but this is unnecessary for temperature regulation. With its high albedo, Venus's current equilibrium temperature is already colder than Earth's. Thus, the planet's surface albedo should be decreased just enough to raise its equilibrium temperature to that of Earth. Then, an Earth-like atmosphere will induce the correct amount of greenhouse warming to produce an Earth-like climate. This will involve lowering Venus's Bond albedo from its current value of 0.76 to 0.62, which could be accomplished by keeping roughly half of the surface mirrored to reflect incoming sunlight. (Some of this mirrored surface would need to be inside colony areas to combat the urban heat island effect.)

Controlling Venus's temperature in this way would result in Earth-like climate patterns without the need for a space-based sunshade, and the greater sun exposure would be beneficial for solar power production and agricultural yields.

\subsection{Rotation}

Venus's slow rotation is often listed as one of the problems of colonizing the planet, with a solar day lasting 117 Earth days. While this is not necessarily an obstacle to a functioning biosphere \citep{Tidal}, a floating artificial surface removes this issue by taking advantage of the superroation of Venus's atmosphere and moving with the prevailing winds. While not as fast as Earth's rotation, if a typical wind speed of $\sim$50 m/s holds at the equator, the artificial surface will rotate roughly once every 9 Earth days. Further study is needed to more accurately predict the interaction of a rigid, floating shell with the superrotating atmosphere.

\subsection{Ecosystems and Infrastructure}

Once floating islands are in place, the next question will be what sorts of natural and urban environments are to be built there. Notably, the full 7440 kg/m$^2$ of lifting capacity must be used to maintain neutral buoyancy. If this much lift is not needed, a thinner surface should be constructed, but it allows the option of having more built-up regions.

Just as on Earth, much of Venus's surface will likely be dedicated to farmland, so we will want to cover most of the surface with soil. One meter of typical arable soil will have a mass of $\sim$2000 kg/m$^2$ \citep{SoilDensity}. Much of this material can be made from regolith mined from the ground. However the water content ($\sim$500 kg/m$^2$) and a seed stock of organic material will need to be imported. Options for sources of water are considered in Section \ref{sec:water}.

The shallowness of this soil layer means that it will be vital to include drainage channels and barriers against soil erosion in the surface design. Hills and valleys could be built into the surface simply by building contours into the honeycomb supporting it, at a negligible cost to lifting capacity. Properly maintained, drainage channels and basins could serve the role of rivers and lakes in the terraformed ecosystem, although they would need to be very shallow, no more than $\sim$8 meters deep.

For urban environments, building single-family homes would pose no problem, nor would smaller high-rise buildings. Typical building regulations for a weight loading of $\sim$500-1000 kg/m$^2$ of floor space in office buildings suggest that they could be built up at least a few floors high, and likely higher because their footprints do not cover all of the land area. If properly supported (e.g. by arches and buttresses in the top layer of the honeycomb, some individual buildings could plausibly be built as tall as the tallest buildings on Earth. Instead, the largest limitation on urban environments is likely to be weight-intensive infrastructure such as subway tunnels and water reservoirs. Thus, construction typical of medium-sized cities will be viable on a terraformed Venus.

\section{Logistics and Time Scales} \label{sec:costs}

\subsection{Energy Requirements}

Any terraforming operation will require very large amounts of energy. These energy requirements may be the limiting factor on terraforming speed. The challenges of industrial production on a planetary scale (e.g. for manufacture of tile aerostats) can be assumed to be solved by automation; self-replicating machines, if used, are more likely to be limited by energy availability than by raw industrial capacity. By estimating the energy requirements of each step in the process, we can compute a minimum timescale for terraforming operations.

Terraforming Venus using the strategy outlined in this paper will require four major, potentially energy-limited operations:
\begin{itemize}
    \item Electrolyzing CO$_2$ into carbon and oxygen.
    \item Separating nitrogen from the atmosphere.
    \item Mining regolith from ground level and delivering it to the artificial surface.
    \item Importing enough water to produce arable soils.
\end{itemize}

CO$_2$ can be electrolyzed at high efficiencies \citep{Electrolysis}. Breaking CO$_2$ into carbon (graphite) and oxygen requires an input of 393.5 kJ/mol, or 8.94 MJ/kg of CO$_2$. In order to process 0.32 bar of CO$_2$ in this way to obtain oxygen and building materials will require $3.33\times10^{10}$ J/m$^2$ of energy flux, or as much sunlight as falls on the planet in 5.8 years. Thus, solar power production will need to proceed on a planetary scale to complete the process on a feasible timescale. With the tile surface already in place, this will be relatively straightforward to implement, but the efficiency of solar power (the likely primary source of energy on Venus) will limit the production rate. With a plausible end-to-end efficiency of 20\%, the oxygen production will take nearly 30 years in total. This of course would require most or all of the initial enclosing surface to be covered with sufficiently lightweight solar panels.

Extracting the nitrogen from the atmosphere will be a much greater task because virtually the whole atmosphere must be processed. The heat of sublimation alone for that much CO$_2$ is more than an order of magnitude greater than that needed to produce the oxygen: 26 kJ/mol, amounting to $5.99\times10^{11}$ J/m$^2$. As a best case scenario, a condenser running at the sublimation point of CO$_2$ at the 1-bar level of 195 K and operating at the Carnot limit would have a coefficient of performance of $\frac{195\,\rm{K}}{288\,\rm{K}-195\,\rm{K}} = 2.06$. Based on the measured temperature-pressure profile \citep{VenusTP}, it may be possible to increase this to $\sim$3 by operating the condenser higher in the atmosphere at $\sim$75 km, but even in this case, the energy requirements to extract all of the nitrogen from the atmosphere are on the order of $2\times10^{11}$ J/m$^2$. With a 20\% efficient source of solar power, this would require all of the solar energy falling on Venus for $\sim$170 years.

It is assumed that the processing of the atmosphere will occur at altitude, where it is needed, to remove the energy costs of lifting that much mass from lower down. (Manually lifting the mass of the atmosphere 75 km from ground level to the condenser would require more energy than the condenser itself.) However, the delivery of soil to the artificial surface \textit{will} require a lifting operation of regolith mined from the ground. The requirement of 1500 kg/m$^2$ of regolith lifted 50 km corresponds to an energy cost of $6.65\times10^8$ J/m$^2$, much less than the other steps in the terraforming operation.

\subsection{Sources of Water} \label{sec:water}

The water content of Venus's atmosphere is 20 ppmv, or 8.6 kg/m$^2$. To produce arable soils over the entire planet would require 500 kg/m$^2$ or $2.30\times 10^{17}$ kg of water, equal to a cube 61.3 km on a side. (This is an order of magnitude greater than the water vapor content of Earth's atmosphere, so it will be sufficient for a complete hydrosphere). This requirement necessitates consideration of how this water is to be delivered.

If the water is delivered before building the tile surface, it would be possible to deliver all of it by colliding a few large comets with the planet. However, this goes against the design philosophy of having as few large, indivisible steps as possible that could hold up the terraforming process, along with facing the problems of safety and security for an operation to deliberately navigate large cometary bodies through the Inner Solar System. During or after construction, water must be delivered either by soft landing or (more likely) by delivering the ice in small pieces that would burn up in the atmosphere a safe distance above the new surface.

With the large amount of mass to be moved, both the fuel and energy costs will be extremely high, and all of the possible sources of water require trade-offs in this area. The inner planets offer greater access to solar energy, which will likely be needed to export water in large quantities. However, exporting water from either Earth or Mars would mean a high $\Delta v$ cost to lift it from the planet's surface. Icy moons in the Outer Solar System suffer similar problems due to the gravity of their host planets. Mining ice from a large centaur like 10199 Chariklo would eliminate the need to lift it out of a planetary gravity well, but it would mean a larger $\Delta v$ cost to move it to the Inner Solar System. Table \ref{tab:deltav} lists the $\Delta v$ costs to export water from various possible sources to Venus. As an average, I will use a (one-way) Hohmann transfer between circular orbits to estimate the cost. (The exact cost will depend on the orbital configuration at the time of launch and may be significantly larger or smaller at different times.) The $\Delta v_{\rm tot}$ column lists the total $\Delta v$ needed to make the delivery with a single impulse, ignoring any more complicated gravity assists.

\begin{table}[htb]
    \centering
    \begin{tabular}{l | r | r | r | r}
    \hline
    Object   & $v_{\rm esc}$ (m/s) & $v_{\rm esc,pl}$ (m/s) & $\Delta v_{\rm Hoh}$ (m/s) & $\Delta v_{\rm tot}$ (m/s) \\
    \hline
    Earth    & 11180 & \      & 2500 & 11460 \\
    Moon     &  2380 &  1440 & 2500 &  3730 \\
    Mars     &  5020 & \      & 4770 &  6920 \\
    Ceres    &   510 & \      & 6380 &  6400 \\
    Callisto &  2440 & 11600 & 6610 & 13570 \\
    Iapetus  &   400 &  4610 & 6020 &  7590 \\
    Chariklo & $<200$ & \     & 5270 &  5270 \\
    Pluto    &  1210 & \      & 3840 &  4026 \\
    Eris     &  1380 & \      & 3090 &  3844 \\
    \hline
    \end{tabular}
    \caption{The $\Delta v$ budget needed to transport water from various solar system bodies to Venus. Because the exact requirements vary with orbital configuration, a one-way Hohman transfer (with uncontrolled reentry) between circular orbits is used to compute an estimated average value.}
    \label{tab:deltav}
\end{table}

Even going far into the Kuiper belt where large amounts of solar energy are not available, the $\Delta v$ costs are significant. Earth's Moon has the lowest $\Delta v$ cost, but the water reserves on the Moon are estimated at only $6\times10^{11}$ kg \citep{MoonWater}, far short of the required amount. Other potential sources of water nearer than the Kuiper belt have a $\Delta v$ cost of at least 5.3 km/s. This would almost certainly require a fuel-to-payload ratio greater than 1, which further compounds the problem of moving large amounts of mass. With such large fuel costs, it may be preferable to build a railgun powerful enough to launch a foil-wrapped projectile to escape velocity, without expending significant fuel. To deliver 500 kg/m$^2$ of water to Venus by this method would require on the order of $10^{10}$ J/m$^2$, albeit not expended on Venus itself. As this would require a large fraction of a planet's energy budget, much like the \textit{in situ} terraforming processes, space-based solar arrays around Earth or Mars may be the only feasible way to produce this much energy.

However, for fast-rotating bodies, a more energy-efficient way to move mass over interplanetary distances would be a space elevator. Unlike either a chemical rocket or a railgun, a space elevator does not suffer significant losses from air resistance and does not need such high power densities. Also, unlike a rotating electromagnetic tether, a space elevator can be powered from the ground and mostly does not need to account for momentum exchange (but see below), allowing it to lift much larger payloads. Most importantly, the energy required to launch a payload via space elevator is only what is needed to lift a payload vertically to the synchronous orbit. Beyond that, centrifugal forces will accelerate the payload outward to the end of the tether, drawing from the planet's rotational energy, giving it large amounts of both rotational and radial velocity.

We can compute the engineering requirements for the space elevator by integrating the apparent gravitational force. In the rotating frame, the apparent gravitational force is given by:
\begin{equation}
    g_{\rm app}(r) = -\frac{GM}{r} + \omega^2r,
\end{equation}

where $\omega$ is the planet's rotation speed. The synchronous orbit occurs where this apparent force is zero. The energy needed to lift the payload from the surface to the synchronous orbit is:
\begin{equation}
    E = \int_{r_s}^{r_{sync}} \left(-\frac{GM}{r^2} + \omega^2r\right)\,dr = GM\left(\frac{1}{r_{sync}} - \frac{1}{r_s}\right) + \frac{\omega^2}{2}(r_{sync}^2 - r_s^2).
\end{equation}

If the far end of the tether is at radius $r_f$, and the payload is able to slide without friction (perhaps by magnetic levitation), then the departure speed will be:
\begin{equation}
    v_d = \left[ 2GM\left(\frac{1}{r_f} - \frac{1}{r_{sync}}\right) + \omega^2(r_f^2 - r_{sync}^2) + (\omega r_f)^2 \right]^{1/2}.
\end{equation}

Solving this expression for $r_f$ and finding the elevator length needed to reach the Hohmann transfer orbit tells us the mechanical requirements for the space elevator. These results are shown for Earth, Mars, and Ceres in Table \ref{tab:elevator}. $\Delta v_{\rm eq}$, the velocity corresponding to an equivalent amount of kinetic energy, is also provided for comparison.

\begin{table}[htb]
    \centering
    \begin{tabular}{l | r | r | r | r}
    \hline
    Object   & E$_{\rm lift}$ (MJ/kg) & $\Delta v_{\rm eq}$ & $r_f$ (km) & $v_{\rm tip}$ (m/s) \\
    \hline
    Earth    & 62.36 & 9840 & 57100 & 4150 \\
    Mars     & 12.58 & 4360 & 53800 & 3810 \\
    Ceres    &  0.13 &  330 & 23500 & 4520 \\
    \hline
    \end{tabular}
    \caption{Energy and design requirements for a space elevator that can lift a payload from Earth, Mars, or Ceres directly to a Hohman transfer orbit to Venus. The $\Delta v$ corresponding to an equal amount of kinetic energy is included for comparison. Ceres is much more efficient, but lacks enough angular momentum to transfer adequate amounts of water to Venus.}
    \label{tab:elevator}
\end{table}

All three options require a similar tip velocity (the speed of the tip of the elevator relative to the planet's rest frame) and thus have similar engineering requirements. These velocities are not attainable with current materials, but are plausibly possible with carbon nanotubes. The tensile stress on a rotating, untapered tether is given by $\sigma = v_{\rm tip}^2\rho/2$. Again using the density of diamond as a worst-case estimate, all of these space elevators have tensile stresses of 25-35 GPa, or likely a bit less because the part of their length that would pass through the planet is not needed. This is within the range reported for carbon nanotubes, and using tapered elevators could reduce this stress significantly.

The energy requirement for this system is much lower for Ceres than for Mars or Earth, which may be useful for rapid delivery of water in the early stages of colonization on Venus. 
The surface of Ceres is estimated to be 5\%-10\% water ice \citep{CeresWater}, so the supply is sufficient. However Ceres is not able to deliver enough water to terraform the whole surface of Venus by this method because it does not have enough angular momentum. The moment of inertia of Ceres is $I_{\rm Ceres} \approx \frac{2}{5}Mr^2 = 8.73\times10^{31}$ kg m$^2$, while the moment of inertia of $2.30\times 10^{17}$ kg of ice lifted to the end of the tether is $I_{\rm ice} = M_{\rm H_2O}r_f^2 = 1.27\times10^{32}$ kg m$^2$. When released, the ice would carry away angular momentum, slowing Ceres's rotation to zero before the process can be completed.

It appears that the majority of the water used to terraform Venus will have to be delivered from either Earth or Mars. Of the two, we may be reluctant to build the planetary infrastructure needed to export water in bulk on Earth (not least because of the environmental impact), and Mars has lower energy costs and less strenuous requirements for a space elevator, so Mars will be the optimal source. (Being much larger than Ceres, the angular momentum exchange would lengthen Mars's day by $\sim$20 seconds.) With the energy requirement to export water from Mars of 12.58 MJ/kg, delivering the requisite water to Venus at 20\% energy efficiency would require all of the solar energy resources available on Mars for $\sim$22 years. This is only a tenth of the investment needed on Venus itself, so it is not a limiting factor in the process. Moreover, water need only be delivered at the rate that the colonized land is built out, which may extend long after the atmospheric engineering is completed. Nonetheless, it is apparent that a large-scale terraforming effort on Mars would likely need to be underway to undertake the task on Venus.

\section{Conclusion} \label{sec:conclusion}

The prospect of terraforming Venus initially seems like an impractical one given the extreme conditions on its surface, the most hostile in the solar system. What proposals have been made face serious logistical challenges given the difficulty of shedding 90 bars of CO$_2$ in a reasonable amount of time and the importation of enough water for conventional oceans. However, the physical similarity of the planet to Earth, its closeness to Earth, and the surprising near-habitable conditions in its upper atmosphere make the planet an enticing target for other forms colonization, particularly ``cloud cities'' that avoid the problems of trying to colonize the surface.

In this paper I lay out a potential program for terraforming Venus that uses the cloud city concept not as an alternative to terraforming, but a stepping stone to it. Expanding the proposed cloud cities to ``cloud continents'' comprising a complete artificial surface could lay the foundation for a terraformed biosphere similar to Earth without the problems of building on the natural ground. A honeycomb of aerostats filled with nitrogen extracted from the atmosphere a few kilometers thick would generate enough lift to support most types of natural and human-made environments, and a breathable atmosphere could be installed above it, with Venus's original atmosphere sequestered below.

This strategy avoids the need to lift Venus's massive atmosphere out of its gravity well or somehow sequester it chemically. The artificial surface could be built with materials that have been demonstrated in the laboratory, and it could support enclosed city-sized colonies before it is completed if so desired. Powered by Venus's abundant solar energy resources, construction could take as little as 200 years in a best-case scenario, similar to other proposals for fast terraforming, with all necessary materials except for water acquired locally. Sufficient water ice to produce arable soils and a functioning hydrosphere could be delivered from Mars via space elevator within the construction time frame. Thus, over the very long term, terraforming Venus could become comparable to terraforming Mars in terms of costs and benefits for human space colonization.

\pagebreak

\section*{Acknowledgements}

This research was supported by an appointment to the NASA Postdoctoral Program at NASA Goddard Space Flight Center, administered by Universities Space Research Association under contract with NASA. I thank Avi Mandell, Michael McElwain, Fred Adams, and Michael Way for helpful conversations.


\end{document}